\documentstyle[prl,eqsecnum,aps,epsfig,floats,tighten]{revtex}
\def\met{\mbox{${\hbox{$E$\kern-0.6em\lower-.1ex\hbox{/}}}_T$}}
\def\lep{{\it l}\hspace{1.5pt}}
\begin{document}
%
%%%%%%%%%%%%%%%%%%%% TITLE PAGE %%%%%%%%%%%%%%%%%%%%%%%%%%%%%%%%%%%%%
%
\title{Search for Second Generation Leptoquark Pairs in 
       $\overline{p}p$ Collisions at $\sqrt s$ = 1.8 TeV \vspace{-0.015cm}}
%
%%% Author List
%
\author{                                                                      
%% names begin here                                                           
B.~Abbott,$^{47}$                                                             
M.~Abolins,$^{44}$                                                            
V.~Abramov,$^{19}$                                                            
B.S.~Acharya,$^{13}$                                                          
D.L.~Adams,$^{54}$                                                            
M.~Adams,$^{30}$                                                              
S.~Ahn,$^{29}$                                                                
V.~Akimov,$^{17}$                                                             
G.A.~Alves,$^{2}$                                                             
N.~Amos,$^{43}$
% ...
\\                                                                            
\vskip 0.05cm                                                                 
\centerline{(D\O\ Collaboration)}                                             
\vskip 0.05cm                                                                 
}                                                                             
\address{                                                                     
\centerline{$^{1}$Universidad de Buenos Aires, Buenos Aires, Argentina}       
\centerline{$^{2}$LAFEX, Centro Brasileiro de Pesquisas F{\'\i}sicas,         
                  Rio de Janeiro, Brazil}                                     
\centerline{$^{3}$Universidade do Estado do Rio de Janeiro,                   
                  Rio de Janeiro, Brazil}                                     
\centerline{$^{4}$Institute of High Energy Physics, Beijing,                  
                  People's Republic of China}                                 
\centerline{$^{5}$Universidad de los Andes, Bogot\'{a}, Colombia}             
\centerline{$^{6}$Universidad San Francisco de Quito, Quito, Ecuador}         
\centerline{$^{7}$Institut des Sciences Nucl\'eaires, IN2P3-CNRS,             
                  Universite de Grenoble 1, Grenoble, France}                 
\centerline{$^{8}$Centre de Physique des Particules de Marseille,             
                  IN2P3-CNRS, Marseille, France}                              
\centerline{$^{9}$LPNHE, Universit\'es Paris VI and VII, IN2P3-CNRS,          
                  Paris, France}                                              
\centerline{$^{10}$DAPNIA/Service de Physique des Particules, CEA, Saclay,    
                  France}                                                     
\centerline{$^{11}$Panjab University, Chandigarh, India}                      
\centerline{$^{12}$Delhi University, Delhi, India}                            
\centerline{$^{13}$Tata Institute of Fundamental Research, Mumbai, India}     
\centerline{$^{14}$Seoul National University, Seoul, Korea}                   
\centerline{$^{15}$CINVESTAV, Mexico City, Mexico}                            
\centerline{$^{16}$Institute of Nuclear Physics, Krak\'ow, Poland}            
\centerline{$^{17}$Institute for Theoretical and Experimental Physics,        
                   Moscow, Russia}                                            
\centerline{$^{18}$Moscow State University, Moscow, Russia}                   
\centerline{$^{19}$Institute for High Energy Physics, Protvino, Russia}       
\centerline{$^{20}$Lancaster University, Lancaster, United Kingdom}           
\centerline{$^{21}$University of Arizona, Tucson, Arizona 85721}              
\centerline{$^{22}$Lawrence Berkeley National Laboratory and University of    
                   California, Berkeley, California 94720}                    
\centerline{$^{23}$University of California, Davis, California 95616}         
\centerline{$^{24}$California State University, Fresno, California 93740}     
\centerline{$^{25}$University of California, Irvine, California 92697}        
\centerline{$^{26}$University of California, Riverside, California 92521}     
\centerline{$^{27}$Florida State University, Tallahassee, Florida 32306}      
\centerline{$^{28}$University of Hawaii, Honolulu, Hawaii 96822}              
\centerline{$^{29}$Fermi National Accelerator Laboratory, Batavia,            
                   Illinois 60510}                                            
\centerline{$^{30}$University of Illinois at Chicago, Chicago,                
                   Illinois 60607}                                            
\centerline{$^{31}$Northern Illinois University, DeKalb, Illinois 60115}      
\centerline{$^{32}$Northwestern University, Evanston, Illinois 60208}         
\centerline{$^{33}$Indiana University, Bloomington, Indiana 47405}            
\centerline{$^{34}$University of Notre Dame, Notre Dame, Indiana 46556}       
\centerline{$^{35}$Purdue University, West Lafayette, Indiana 47907}          
\centerline{$^{36}$Iowa State University, Ames, Iowa 50011}                   
\centerline{$^{37}$University of Kansas, Lawrence, Kansas 66045}              
\centerline{$^{38}$Kansas State University, Manhattan, Kansas 66506}          
\centerline{$^{39}$Louisiana Tech University, Ruston, Louisiana 71272}        
\centerline{$^{40}$University of Maryland, College Park, Maryland 20742}      
\centerline{$^{41}$Boston University, Boston, Massachusetts 02215}            
\centerline{$^{42}$Northeastern University, Boston, Massachusetts 02115}      
\centerline{$^{43}$University of Michigan, Ann Arbor, Michigan 48109}         
\centerline{$^{44}$Michigan State University, East Lansing, Michigan 48824}   
\centerline{$^{45}$University of Nebraska, Lincoln, Nebraska 68588}           
\centerline{$^{46}$Columbia University, New York, New York 10027}             
\centerline{$^{47}$New York University, New York, New York 10003}             
\centerline{$^{48}$University of Rochester, Rochester, New York 14627}        
\centerline{$^{49}$State University of New York, Stony Brook,                 
                   New York 11794}                                            
\centerline{$^{50}$Brookhaven National Laboratory, Upton, New York 11973}     
\centerline{$^{51}$Langston University, Langston, Oklahoma 73050}             
\centerline{$^{52}$University of Oklahoma, Norman, Oklahoma 73019}            
\centerline{$^{53}$Brown University, Providence, Rhode Island 02912}          
\centerline{$^{54}$University of Texas, Arlington, Texas 76019}               
\centerline{$^{55}$Texas A\&M University, College Station, Texas 77843}       
\centerline{$^{56}$Rice University, Houston, Texas 77005}                     
\centerline{$^{57}$University of Washington, Seattle, Washington 98195}       
}                                                                             
%
%%% End of author/institution list.
%
\maketitle
\begin{abstract}
  We have searched for second generation leptoquark (LQ) pairs in 
  the $\mu\mu$+jets channel using 94~$\pm$~5~pb$^{-1}$ of
  $\overline{p}p$ collider  data collected by the D\O\ experiment at
  the Fermilab Tevatron during 1993--1996. No evidence for a signal is 
  observed. These results are combined with those from the
  $\mu\nu$+jets and $\nu\nu$+jets channels to obtain 95\% confidence
  level (C.L.) upper limits on the LQ pair production cross section as
  a function of mass and $\beta$, the branching fraction of a LQ
  decay into a charged lepton and a quark. Lower limits of
  200(180)~GeV/$c^2$ for $\beta=1$(${1\over2}$) are set at the 95\%\
  C.L. on the mass of scalar LQ. Mass limits are also set on vector
  leptoquarks as a function of $\beta$.
\end{abstract}
\twocolumn
%
%%%%%%%%%%%%%%%%%% Introduction %%%%%%%%%%%%%%%%%%%%%%%%%%%%%%%%%%%%%
%
  The observed symmetry in the spectrum of fundamental particles
  between leptons (\lep) and quarks ($q$) has led to suggestions 
  of the existence of leptoquarks (LQ)\cite{generic_lq}. Leptoquarks
  would carry both lepton and quark quantum numbers, and would decay
  to \lep $q$ systems. Although, in principle, leptoquarks could decay 
  to any \lep $q$ combinations, limits on flavor-changing neutral
  currents, rare lepton-family violating decays, and proton decay,
  suggest that leptoquarks would couple only within a single
  generation\cite{fcnc}. This implies the existence of three LQ
  generations, analogous to the fermion generations in the standard
  model.{\par}

  At the Fermilab Tevatron, leptoquarks are predicted\cite{pair_pro} to
  be produced dominantly via gluon ($g$) splitting, \mbox{$\overline{p}p
  \to g + X \to LQ\overline{LQ} + X$}.  This Letter reports on an
  enhanced search for second generation lepto\-quark pairs produced in
  $\overline{p}p$ interactions at a center-of-mass energy
  \mbox{$\sqrt{s}$ = 1.8~TeV}. The experimental signature considered
  is when both leptoquarks decay via \mbox{LQ $\to \mu q$}, where
  $q$ can be either a strange or a charm quark depending on the electric 
  charge of the LQ. The corresponding
  experimental cross section is \mbox{$\beta^2 \times
  \sigma(\overline{p}p \to LQ\overline{LQ})$}, where $\beta$ is the
  unknown branching fraction of a LQ to a muon ($\mu$) and a
  \mbox{quark (jet)}.{\par}

  Previous studies by the D\O\cite{D02GenLQ} and CDF\cite{CDF2GenLQ}
  collaborations have considered pair production of scalar leptoquarks
  in $\mu\mu$+jets final states. These studies provide lower limits on 
  the mass of LQs of 119~GeV/$c^2$ and 202~GeV/$c^2$, respectively,
  for $\beta$ = 1. Lower limits of 160~GeV/$c^2$ for $\beta$ = 1/2
  were obtained by D\O\ from the $\mu\nu$+jets final
  state\cite{munulq} and by CDF from the $\mu\mu$+jets final
  state\cite{CDF2GenLQ}. For $\beta$ = 0, D\O\ has obtained a lower
  limit of 79~GeV/$c^2$ from the $\nu\nu$+jets
  channel\cite{slqnunu}.{\par}

  The present study is complementary to
  previous D\O\ searches in the $\mu\nu$+jets\cite{munulq} and
  $\nu\nu$+jets\cite{slqnunu} final states, and greatly extends the 
  previous search in the $\mu\mu$+jets channel\cite{D02GenLQ}. 
  The sensitivity for detection of leptoquarks is increased by considering 
  a larger data set that uses the calorimeters to identify muon
  candidates, and employs several optimization techniques to enhance
  efficiency. These results are combined with results from
  other decay channels to improve mass limits on LQs. (A detailed
  description of this analysis can be found in Ref.\cite{mywork}.){\par}
%
%%%%%%%%%%%%%%%%%% The D0 Detector %%%%%%%%%%%%%%%%%%%%%%%%%%%%%%%%%%
%
  The D\O\ detector\cite{d0nim} consists of three major components: 
  an inner detector for tracking charged particles, a uranium/liquid 
  argon calorimeter for measuring electromagnetic and hadronic 
  showers, and a muon spectrometer consisting of magnetized iron 
  toroids and three layers of drift tubes. Jets are measured with an 
  energy resolution of approximately $\sigma (E)/E$ = 0.8/$\sqrt{E}$ 
  ($E$ in GeV). Muons are measured with a momentum resolution of 
  $\sigma (1/p) = 0.18 (p-2)/p^2 \oplus 0.003$ ($p$ in GeV/$c$).{\par}

  Event samples are obtained from triggers requiring the presence of a 
  muon candidate with transverse momentum $p_T^{\mu} >$ 5~GeV/$c$ in 
  the fiducial region \mbox{$|\eta_{\mu}| < 1.7$ ($\eta \equiv 
  -\ln[\tan ({1\over2}\theta)]$}, where $\theta$ is the polar angle 
  of a track with respect to the $z$--axis taken along the direction
  of the proton beam), and at least one jet candidate with transverse
  energy $E_T^j >$ 8~GeV and $|\eta_j| <$ 2.5. The data correspond 
  to an integrated luminosity of 94~$\pm$~5~pb$^{-1}$ collected during 
  the 1993--1995 and 1996 Tevatron collider runs at Fermilab
  \cite{lumblurb}.{\par}

%
%%% Off line particle ID %%%%%%%%%%%%%%%%%%%%%%%%%%%%%%%%%%%%%%%%%%%
%  
  Jets are measured in the calorimeters and are reconstructed offline 
  with a cone algorithm having radius 
  \mbox{${\cal R} \equiv \sqrt{\Delta\phi^2 + \Delta\eta^2}$ = 0.5}. In 
  the final event sample, \mbox{two} or more jets are required with 
  \mbox{$E_T^j >  20$~GeV} within \mbox{$|\eta_j|<3.0$}.{\par}

  Muon candidates reconstructed in the muon spectrometer are required
  to have a track that projects back to the interaction vertex. 
  The track is required to be consistent with a muon of 
  \mbox{$p_T^{\mu}>$ 20~GeV/$c$} and $|\eta_{\mu}|<1.7$. In addition,
  the muon is required to deposit energy in the calorimeter
  consistent with the passage of a minimum ionizing particle (MIP). To
  reduce backgrounds from heavy quark production, candidate muons are 
  required to be isolated from all jets passing the selection criteria
  listed above by $\Delta R_{\mu j} > $ 0.5 in the $\eta-\phi$ plane.{\par}
  
  Single muon candidates can also be tracked in the calorimeters,
  where an isolated high--$p_T$ muon deposits only a small fraction of
  its total energy. This results in a unique energy signature
  consisting of energy from a MIP ($E_{\rm MIP}$)\cite{munulq,zgamma} and
  a large transverse energy imbalance (\met) in the calorimeter that is
  proportional to the muon momentum, and points in the azimuthal
  direction of the $E_{\rm MIP}$. Muon candidates are restricted to the region
  $|\eta| < 1.7$, and are required to have 
  \mbox{$|\Delta\phi$($E_{\rm MIP}$--\met)$| < 0.25$ radians}. The kinematic 
  quantities (e.g., $p_T^{\mu}$) of
  these candidates are calculated using the ($\eta, \phi$) direction
  of the $E_{\rm MIP}$ and the component of the \met\ along the azimuthal
  direction of the $E_{\rm MIP}$.{\par} 

  Dimuon candidate events are required to have two muons with
  $p_T^{\mu} > $ 20~GeV/$c$. At least one muon must be in the central
  muon spectrometer ($|\eta_{\mu}|<1.0$). A second muon with
  $|\eta_{\mu}|<1.7$ may be identified using either the muon
  spectrometer or the calorimeters.{\par}

  After obtaining a sample of $\mu\mu$+jets events, a selection is 
  applied to the event topology.  Heavy LQ pairs are expected to have
  a smaller Lorentz boost, and to decay more symmetrically, than the
  background events. To take advantage of these differences, the 
  sphericity in the center-of-mass frame (${\cal{S}}_{{\rm
  CM}}$) is required to be greater than 0.05. ${\cal{S}}_{{\rm CM}}$
  is defined as $1.5 (\lambda_1 + \lambda_2)$, with $\lambda_1 \le
  \lambda_2 \le \lambda_3$ being the normalized eigenvalues of the
  momentum tensor. The momentum tensor is formed from the $E_T$
  ($p_T$) of all jets (muons) in an event, and ${\cal{S}}_{{\rm CM}}=$
  0 (1) corresponds to a linear (spherical) topology.{\par}
  \begin{figure}[t]
  \vbox{
  \vspace{-0.40in}
  \centerline{
  \psfig{figure=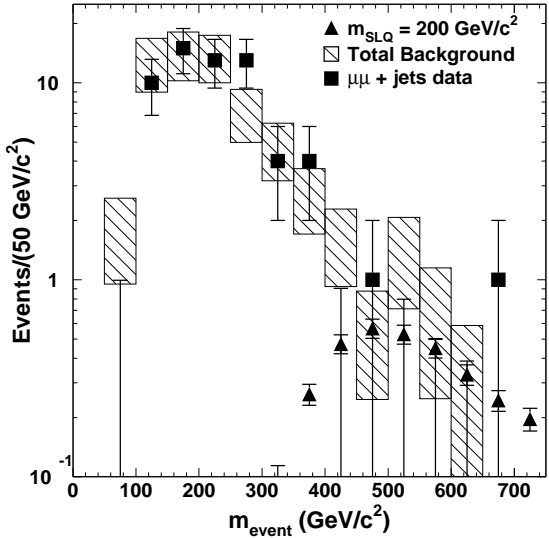,width=3.0in}}
  \caption{Invariant mass of $\mu\mu$+jets events. The mass is 
           calculated from all muons and jets that pass the selection 
           criteria. The hatched regions give the background
           estimation, the square points are the $\mu\mu$+jets data,
           and the triangular points are the prediction for $S_{{\rm
           LQ}}$ from the Monte Carlo. Uncertainties on bins with no 
	   data points are obtained from the 68\%\ confidence interval.}
  \label{fig:fig1}
  }
  \vspace{-15pt}
  \end{figure}
%
%%%%%%%%%%%%%%%%%%%%%%%%%%%%% Monte Carlo %%%%%%%%%%%%%%%%%%%%%%%%%%
%
  Leptoquark events are simulated with the {\footnotesize
  {ISAJET}}\cite{isajet} Monte Carlo event generator for scalar 
  LQ ($S_{{\rm LQ}}$), and with {\footnotesize{PYTHIA}}\cite{pythia}
  for vector LQ ($V_{LQ}$). The detection efficiencies for $S_{{\rm
  LQ}}$ and $V_{{\rm LQ}}$ of the same mass are found to be consistent
  within the uncertainties. For massive vector leptoquarks
  ($m_{V_{LQ}} >$ 200~GeV/$c^2$), efficiencies are insensitive to
  differences between minimal vector (MV, \mbox{$\kappa_G=1$},
  \mbox{$\lambda_G=0$}\cite{VLQCouple}) and Yang-Mills (YM,
  \mbox{$\kappa_G=\lambda_G=0$}\cite{VLQCouple}) couplings to standard
  model bosons\cite{1GenVLQ}. Consequently, the $S_{\rm LQ}$ Monte
  Carlo is used to represent the shapes of distributions for both
  $S_{\rm LQ}$ and $V_{\rm LQ}$ analyses.{\par} 

  The leptoquark cross sections for $S_{{\rm LQ}}$ are next-to-leading-order 
  calculations (NLO)\cite{kraemer} at a renormalization scale 
  \mbox{$\mu = m_{S_{{\rm LQ}}}$}. The uncertainties are determined from 
  variation of the renormalization/factorization scale from $2 
  m_{S_{{\rm LQ}}}$ to ${1\over2} m_{S_{{\rm LQ}}}$. Both types of 
  $V_{{\rm LQ}}$ cross sections are calculated to leading-order (LO) 
  at $\mu=m_{V_{{\rm LQ}}}$\cite{VLQCouple}.{\par}

  The dominant backgrounds are due to $W$+jets and $Z$+jets production, 
  and are simulated using {\footnotesize{VECBOS}} \cite{vecbos} at the 
  parton level and {\footnotesize{HERWIG}} \cite{herwig} for
  parton fragmentation. Background due to $WW$ production is 
  simulated with {\footnotesize{PYTHIA}}\cite{pythia}. Background from 
  $t\overline{t}$ production is simulated using
  {\footnotesize{HERWIG}} with a top quark mass of 170~GeV/$c^2$. 
  All Monte Carlo samples are processed through a detector simulation
  program based on the {\footnotesize{GEANT}}\cite{geant} package.{\par}
%
%%%%%%%%%%%%%%%%%%%%%%%%%%%%% DiMuon analysis %%%%%%%%%%%%%%%%%%%%%%%
%
  After initial selection, there are 53 events in the data sample 
  consistent with an estimated background of 53$\pm$13 events. The
  distribution in invariant mass ($m_{\text{event}}$) calculated from
  all muons and jets passing the selection criteria is given in
  Fig.~1. The largest expected background is from $W$+jets (43$\pm$13 
  events) where \met\ from a neutrino is misidentified as a second
  muon when low-energy jets or calorimeter noise mimic the energy
  signature of a MIP. The other backgrounds are from $Z$+jets events
  (5.6$\pm$0.9), $WW$ events (2.3$\pm$0.9, consistent with previous
  experimental limits at D\O\cite{WWgamma}), and $t\overline{t}$
  events (2.1$\pm$0.6). The uncertainty in the background estimate is
  dominated by the statistical uncertainty of the $W$+jets Monte Carlo
  and the systematic uncertainty in the $W$+jets production cross
  section. The estimate for the production of 200~GeV/$c^2$ scalar
  leptoquarks that pass all of the previous selection requirements is 
  3.7$\pm$0.4 events. All leptoquark production estimates are for
  200~GeV/$c^2$ $S_{{\rm LQ}}$, and use the NLO cross section at a
  scale \mbox{$\mu = 2m_{S_{{\rm LQ}}}$}.{\par}
  
  A neural network (NN) analysis\cite{nn} is employed to separate any
  possible signal from background. The \mbox{NN} is trained using a mixture
  of $W$+jets, $Z$+jets, and $t\overline{t}$ background Monte Carlo
  events, and an independently generated $S_{{\rm LQ}}$ Monte Carlo
  sample for a mass $m_{S_{{\rm LQ}}} = $~200~GeV/$c^2$. The \mbox{NN} uses
  seven inputs: [$E_T^{j_1}$, $E_T^{j_2}$, $p_T^{{\mu}_1}$,
  $p_T^{{\mu}_2}$, ($E_T^{j_1}+E_T^{j_2}$), $m_{\text{event}}$ and 
  $ (E_T^{j_1}+E_T^{j_2})/{\sum}E_T^{j_i}$, where jets (muons) are
  ordered in $E_T$ ($p_T$)], and 15 \mbox{nodes} in a single hidden layer to
  calculate an output. The network output ($D_{NN}$) is shown in Fig.~2.{\par}
  
  \begin{figure}[t]
  \vbox{
  \vspace{-0.40in}
  \centerline{
  \psfig{figure=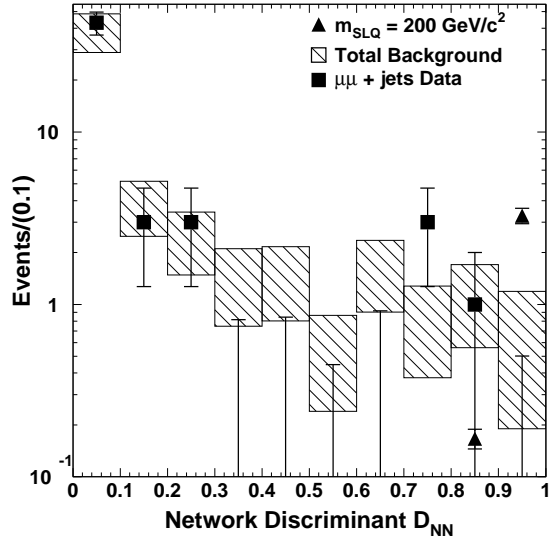,width=3.0in}}
  \caption{Output of the neural network. The network calculates a 
           value for each event based on the inputs (see text) and a 
           set of internal values which are determined during network 
           training on $S_{{\rm LQ}}$ and background Monte Carlo.}
  \label{fig:fig2}
  }
  \vspace{-15pt}
  \end{figure}
  
  No evidence of a signal is seen either in the $D_{NN}$ discriminant
  or in any kinematic distribution. The $D_{NN}$ selection 
  is optimized for the calculation of limits using a measure of
  sensitivity\cite{munulq}  calculated from samples of $S_{{\rm LQ}}$
  and background Monte Carlo. The requirement is set at
  $D_{NN}>0.9$. For this selection no events are observed, consistent
  with an estimated background of $0.7{\pm}0.5$ events (0.49$\pm$0.16 
  $t\overline{t}$, 0.15$\pm$0.04 $Z$+jets, 0.05$\pm$0.05 $WW$, and
  $0^{+0.5}_{-0.0}$ $W$+jets events). The estimate for 200~GeV/$c^2$ 
  $S_{{\rm LQ}}$ production is $3.3{\pm}0.3$ events.{\par}
  \begin{table}
  \vspace{-20pt}
  \begin{tabular} { c c c c c c c }
    LQ Mass  & Efficiency & 
    ${\sigma}_{\mu\mu+{\rm jets}}^{95\%}$   &
    ${\sigma}_{{\rm combined}}^{95\%}$   & 
    ${\sigma}_{S_{{\rm LQ}}}$ & 
    ${\sigma}_{{\rm MV}}    $ &
    ${\sigma}_{{\rm YM}}    $  \\
    (GeV/$c^2$) & (\%) & (pb) & (pb) & (pb) & (pb) & (pb) \\
  \hline
  140 & 10.3$\pm$0.3$\pm$1.1 & 0.33 & 0.55 & 1.5  & 20   & 100 \\
  160 & 14.5$\pm$0.3$\pm$1.6 & 0.24 & 0.38 & 0.68 & 8.0  & 50  \\
  180 & 18.9$\pm$0.4$\pm$2.1 & 0.18 & 0.31 & 0.32 & 4.0  & 20  \\
  200 & 21.8$\pm$0.4$\pm$2.1 & 0.16 & 0.26 & 0.16 & 2.0  & 10  \\
  220 & 22.6$\pm$0.4$\pm$2.4 & 0.15 & 0.26 & 0.08 & 0.90 & 5.0 \\
  240 & 23.5$\pm$0.4$\pm$2.5 & 0.15 & 0.24 & 0.04 & 0.45 & 2.5 \\
  260 & 24.3$\pm$0.5$\pm$2.6 & 0.15 & 0.24 & 0.02 & 0.25 & 1.2 \\
  280 & 26.0$\pm$0.5$\pm$2.8 & 0.13 & 0.22 &      & 0.12 & 0.60\\
  300 & 25.3$\pm$0.5$\pm$2.7 & 0.13 & 0.23 &      & 0.06 & 0.35\\
  350 & 25.7$\pm$0.5$\pm$2.8 & 0.13 & 0.23 &      &      & 0.06\\
  400 & 25.7$\pm$0.5$\pm$2.8 & 0.13 & 0.22 &      &      &     
  \end{tabular}
  \vspace{0.05in}
  \caption{Leptoquark detection efficiencies (with statistical and 
           systematic uncertainties) and 95\%\ C.L. cross section
           limits for leptoquarks in the $\mu\mu$+jets channel and for
           the combination of all decay channels at
           $\beta={1\over2}$. Cross sections for $S_{{\rm LQ}}$ (NLO)
           and $V_{{\rm LQ}}$ (LO) pair production are also shown.} 
  \label{tab:table1}
  \end{table}

  \begin{figure}[ht]
  \vbox{
  \vspace{-20pt}
  \centerline{
  \psfig{figure=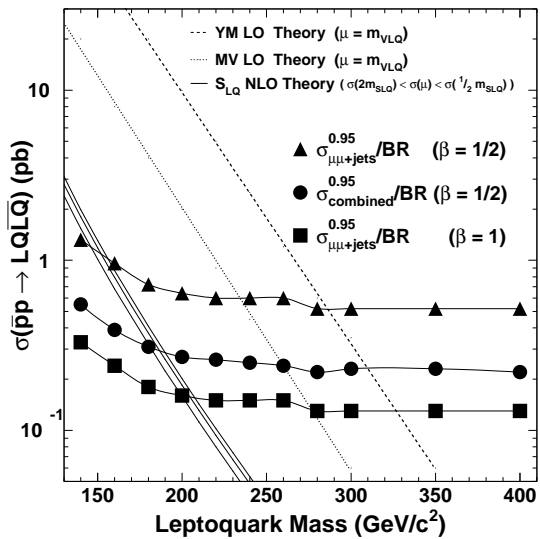,width=3.0in}}
  \caption{95\%\ C.L. limits on pair production cross
           sections. Results are shown for the $\mu\mu$+jets channel 
           ($\sigma_{\mu\mu+{\rm jets}}^{0.95}$) for $\beta = 1,
           {1\over2}$, and for all combined searches ($\sigma_{\rm
           combined}^{0.95}$) at $\beta ={1\over2}$.}
  \label{fig:fig3}
  }
  \vspace{-10pt}
  \end{figure}
  The selection criteria are applied to the Monte Carlo for a 
  range of LQ masses. The leptoquark detection efficiencies, estimated
  to be 10\%\--26\%\, depending on the LQ mass, are listed in
  Table~I, along with the 95\%\ confidence level (C.L.) upper
  limits on the cross sections. The limits are calculated using a
  Bayesian approach, with a flat prior distribution for the signal
  cross section. The \mbox{statistical} and systematic uncertainties on the
  efficiencies, the integrated luminosity (5\%), and the background
  estimate are included in the calculation assuming Gaussian prior
  distributions. It should be noted that the cross section limits for
  the $\mu\mu$+jets channel are independent of $\beta$, which enters
  only when comparing experimental limits with theory. A particular
  $\beta$ is given for the combined result since that value determines
  the relative contribution of each channel to the total cross
  section.{\par}

  The dominant (10\%) systematic uncertainty in the efficiencies is
  due to uncertainty in the simulation. In addition, there are 
  approximately equal uncertainties in the jet energy
  scale\cite{jetenergy} and the trigger efficiency/spectrometer
  resolution for high-$p_T$ muons (6.6\%\ and 6.4\%\,
  respectively).{\par} 

  Figure~3 shows the limits on the pair production cross sections for
  scalar and vector leptoquarks obtained from this search, corrected
  for the branching ratio (BR = $\beta^2$ for $\mu\mu$+jets). The
  results are given for $\beta$ = 1 and $1\over2$. The lower mass
  limits at the 95\%\ confidence level obtained from comparing the
  cross section limits with the theory cross sections at $\mu = 2
  m_{S_{LQ}}$ for the $\mu\mu$+jets
  decay channel at ${\beta}=1$ $(1/2)$ are: 200 (145)~GeV/$c^2$, 270
  (225)~GeV/$c^2$ and 325 (280)~GeV/$c^2$ for scalar, MV, and YM
  vector couplings, respectively.{\par}
  \begin{figure}[t]
  \vbox{
  \vspace{-0.40in}
  \centerline{
  \psfig{figure=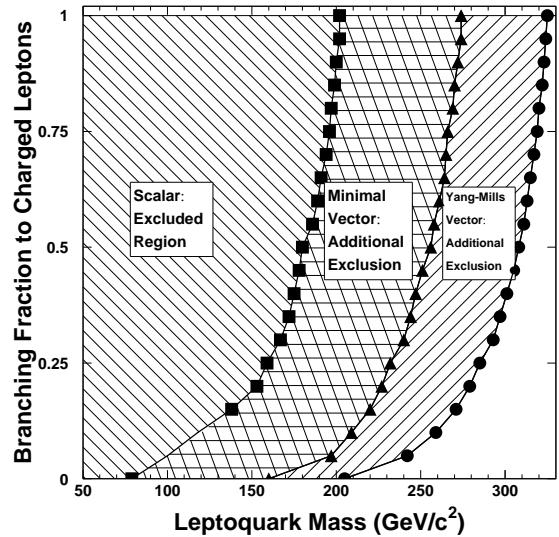,width=3.0in}}
  \caption{The regions in the $\beta-m_{{\rm LQ}}$ plane excluded by 
           combining the results of the $\mu\mu$+jets, $\mu\nu$+jets,  
           and $\nu\nu$+jets searches. The area to the left of each 
           curve is excluded for that type of coupling, at the 95\%\
           confidence level.}
  \label{fig:fig4}
  }
  \vspace{-5pt}
  \end{figure}

  \begin{table}[t]
  \begin{tabular} { c c c c }
    $\beta$ & Scalar (GeV/$c^2$) & MV (GeV/$c^2$) & YM (GeV/$c^2$) \\
  \hline

  $1$   &  $200$&  $275$ & $325$ \\
  $1/2$ &  $180$&  $260$ & $310$ \\
  $0$   &  $79 $&  $160$ & $205$

  \end{tabular}
  \caption{Combined 95\%\ C.L. lower mass limits for second generation
  leptoquarks.}
  \vspace{-0.2in}
  \label{tab:table2}
  \end{table}
  The results from the $\mu\mu$+jets (BR = $\beta^2$) search are
  combined with results from previous second generation leptoquark
  searches in the $\mu\nu$+jets (BR = 2$\beta (1-\beta)$)\cite{munulq}
  and $\nu\nu + {\rm jets}$ (BR = $ (1-\beta)^2$)\cite{slqnunu} 
  channels. Limits on the combined cross section (BR = 1) are listed
  in Table~I, for $\beta$ = 1/2. These limits are also shown in
  Fig.~3, and the lower mass limits obtained are: 180~GeV/$c^2$
  ($S_{{\rm LQ}}$), 260~GeV/$c^2$ (MV), and 310~GeV/$c^2$ (YM), all at 
  the 95\%\ confidence level. Mass limits calculated from the combination of
  channels as a function of $\beta$ are shown in Fig.~4 and summarized
  in Table~II.{\par} 

  In conclusion, a search has been performed for second generation 
  leptoquark pairs decaying via LQ $\to \mu q$ using
  94~$\pm$~5~pb$^{-1}$ of data. No evidence is found for a signal, and limits are
  set at the 95\%\ confidence level on the mass of second generation 
  leptoquarks. By combining these results with those from 
  previous studies comprehensive limits on second generation leptoquarks 
  are obtained. These are shown as exclusion contours 
  constraining the possible values of $\beta$ and $m_{{\rm LQ}}$ by
  coupling.{\par}
%
%%%%%%%%%%%%%%%%%%% Acknowledgement paragraph %%%%%%%%%%%%%%%%%%%%%%%
%
We thank the Fermilab and collaborating institution staffs for 
contributions to this work, and acknowledge support from the 
Department of Energy and National Science Foundation (USA),  
Commissariat  \` a L'Energie Atomique (France), 
Ministry for Science and Technology and Ministry for Atomic 
   Energy (Russia),
CAPES and CNPq (Brazil),
Departments of Atomic Energy and Science and Education (India),
Colciencias (Colombia),
CONACyT (Mexico),
Ministry of Education and KOSEF (Korea),
CONICET and UBACyT (Argentina),
and the A.P. Sloan Foundation.
%
%%%%%%%%%%%%%%%%%%% References %%%%%%%%%%%%%%%%%%%%%%%%%%%%%%%%%%%%%%
%

%
%%% End of Document
%
\end{document}